# Electrical Control of Quantum Emitters in a Van der Waals Heterostructure

Simon J. U. White[1,#], Tieshan Yang[1,2,#], Nikolai Dontschuk[3], Chi Li[1], Zai-Quan Xu[1], Mehran Kianinia[1,2], Alastair Stacey[4], Milos Toth[1,2] and Igor Aharonovich[1,2]

1. School of Mathematical and Physical Sciences, University of Technology Sydney, Ultimo, New South Wales 2007, Australia

2. ARC Centre of Excellence for Transformative Meta-Optical Systems, University of Technology Sydney, Ultimo, New South Wales 2007, Australia

3. School of Physics, University of Melbourne, Parkville, Victoria 3010, Australia

4. School of Science, RMIT University, Melbourne, Victoria 3001, Australia

# These authors contributed equally

*Controlling and manipulating individual quantum systems in solids underpins the growing interest in development of scalable quantum technologies. Recently, hexagonal boron nitride (hBN) has garnered significant attention in quantum photonic applications due to its ability to host optically stable quantum emitters. However, the large band gap of hBN and the lack of efficient doping inhibits electrical triggering and limits opportunities to study electrical control of emitters. Here, we show an approach to electrically modulate quantum emitters in n hBN–graphene van der Waals heterostructure. We show that quantum emitters in hBN can be reversibly activated and modulated by applying a bias across the device. Notably, a significant number of quantum emitters are intrinsically dark, and become optically active at non-zero voltages. To explain the results, we provide a heuristic electrostatic model of this unique behaviour. Finally, employing these devices we demonstrate a nearly-coherent source with linewidths of ~ 160 MHz. Our results enhance the potential of hBN for tuneable solid state quantum emitters for the growing field of quantum information science.*

Van der Waals (vdW) heterostructures have emerged as a fascinating platform to study light-matter interaction at the nanoscale[1-11]. Assembling various atomically thin crystals has enabled the observation of new physical phenomena in these unconventional materials, including superconductivity[12], interlayer excitons[13], moire lattices[3, 14] and correlated electronic systems[15]. Furthermore, advanced practical devices such as broadband photodetectors, efficient light emitting diodes and nanoscale lasers have also been realized from a variety of vdW crystals[16]. Indeed, control over light emission from a selected family of transition metal di-chalcogenides enabled optical detection of valley states, and observation of exciton-polariton condensates even at room temperature[17-20].

Of particular interest is the ability to manipulate light emission from single point defects, commonly referred to as single photon emitters (SPEs), as they are critical building blocks for quantum technologies[1]. Hexagonal boron nitride (hBN), a wide band gap vdW crystal, has been extensively studied in recent years as a vdW host of SPEs that are ultra bright and optically stable. In addition, hBN SPEs exhibit spin – photon interface and can be engineered on demand in an atomically thin crystal[21, 22]. This combination of photophysical properties foreshadows ample opportunities for their utilisation as quantum sources and quantum repeaters in scalable quantum photonic devices. An outstanding challenge for solid state SPEs is to realize electrical control of the optical emission. This challenge stems from the fact that most hosts of defect-based SPEs are wide band gap materials in which p-type or n-type doping is limited[23, 24]. Indeed, even for well-studied materials such as diamond or silicon carbide, electrical modulation of quantum emitters is limited to specific defects and often requires cumbersome device engineering[25-28].

Here we demonstrate a facile and scalable approach to electrically modulate quantum emitters in hBN – graphene heterostructures. Our experiments show that SPEs in hBN can be controllably activated and modulated by applying a voltage across the devices. Intriguingly, we show that most of the quantum emitters become optically active at non-zero voltages, in contrast to what has been observed in the case of defects in 3D crystals. We interpret our results in the context of electrically-induced changes in the charge states of the hBN defects and provide electrostatic models to support the experimental findings.

Figure 1a is a schematic illustration of the heterostructure devices used in this study. The device structure consists of multilayer graphene (MLG), a hBN capping layer and a hBN emitter layer stacked vertically on p-type silicon with a 285 nm thermal oxide. Bias is applied between the bottom p-type silicon and MLG. An optical image of the device is shown in Figure 1b. The black, light blue and green dashed lines indicate the boundaries of the MLG, the hBN capping layer and the hBN layer that hosts the quantum emitters, respectively. The capping layer (~20 nm) is used to prevent quenching of emitters in the active hBN layer by MLG. Details of the fabrication process can be found in the methods section.

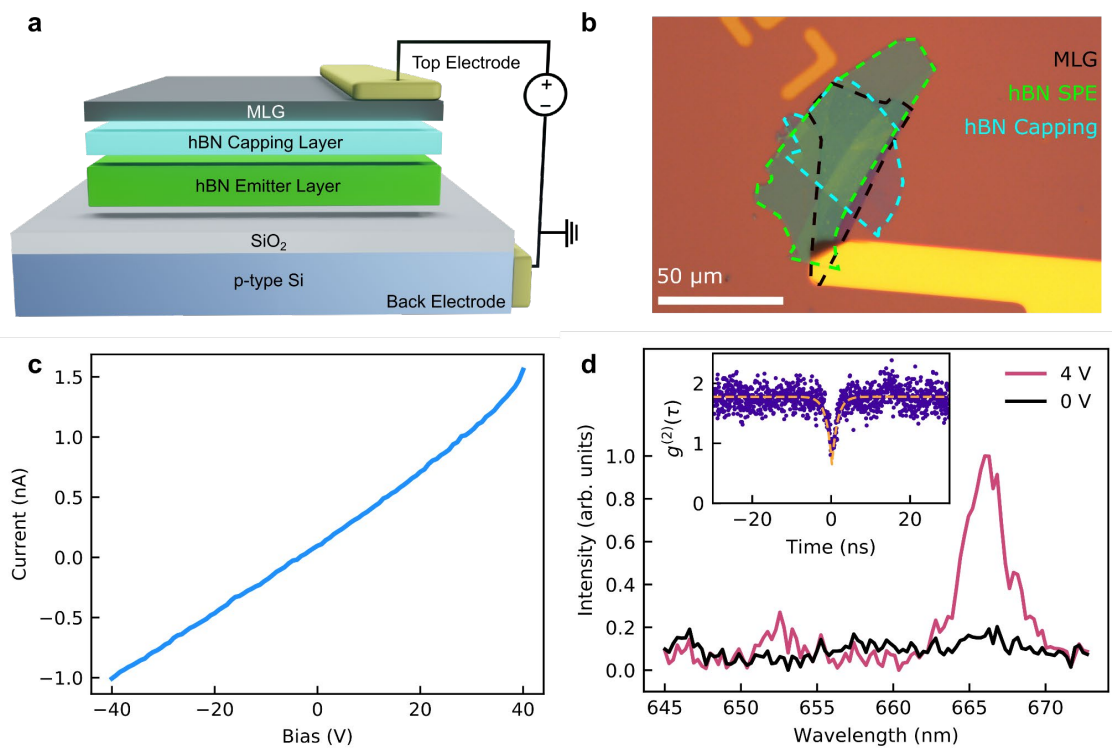

*Figure 1: The hBN/MLG heterostructure device. **a**. Schematic of the device and its operating principle. The device is biased using gold electrodes in contact with MLG and p-type silicon. **b**. Optical image of a device consisting of MLG, a hBN capping layer and a hBN layer that contains SPEs. The substrate is p-type silicon with a 285 nm thermal oxide layer. Each layer is outlined by dashed lines. **c**. I-V cure from the device showing a negligible leakage current. **d**. Normalized PL spectra of an emitter*

*in the heterostructure at a bias of 0 (black) and 4 V (purple), respectively. Inset: second autocorrelation function measured from the emitter at 4 V, confirming its a single photon source.*

To characterise the device, we first measured a current-voltage (I-V) curve by sweeping the bias from -40 V to 40 V. The current scales linearly with voltage, as shown in Figure 3c, and the maximum measured current is less than 1.5 nA. This is an upper bound on the current through the hBN layers since the top electrode is in contact with both the MLG and the oxide layer (see Figure 1b). The I-V curve shows that the device behaves as a capacitor that generates an electric field within the hBN layers. Additional electrical measurements from the device are included in the supporting information.

Next, we study the optical properties of quantum emitters embedded in the heterostructure. All optical measurements were performed using a 532 nm continuous wave excitation laser, and a custom-built confocal microscope (see methods for details). Figure 1d shows photoluminescence (PL) spectra from one emitter at ambient conditions, using a bias of 0 (black curve) and 4 V (purple curve). Remarkably, a clear peak at 666 nm arises when the voltage is switched from 0 to 4 V, indicating activation of the emitter by the applied bias. A second-order autocorrelation function measured from this emitter is shown in the inset of Figure 1d. The dip at zero delay time confirms that the defect is a single photon emitter. Additional spectra of emitters are presented in the SI.

To elucidate this peculiar behaviour, the hBN/MLG heterostructure device was loaded into a closed cycle He cryostat operating at 4 K. Electrical control of the hBN quantum emitters is shown in Figure 2. The bias-dependence of PL spectra from two emitters are plotted in Figure 2a and 2b. The spectra are normalised for clarity, and illustrate two distinct behaviours observed predominantly at under positive (Figure 2a) and negative (Figure 2b) bias applied to the MLG electrode. The emitter in Figure 2a does not fluoresce at zero bias. However, as the bias is increased, the emitter becomes active at ~8 V, the brightness increases up to ~15 V where it goes through a maximum and decreases as the bias is increased further. It becomes inactive at ~ 22 V, and is not restored within our experimental conditions. On the other hand, the emitter in Figure 2b shows a completely different behavior. As the bias is reduced from 0V to - 30V, the emission intensity increases gradually, and remains optically active even under -30V. This is unexpected, given that under positive biases, there was only a window of voltages under which the emission was persistent. This would be explained later in detail. Note, that in both cases, a minor shift of the emission was observed, as expected, due to the stark shift. The direction of the Stark shifts depends on the polarity of the applied bias and the dipole orientation of each emitter.

The emission intensity can be further tuned dynamically with the applied bias. This is shown in Figure 2c, where an emitter is modulated using a square wave voltage function oscillating between 0 and +10 V. The period of the intensity resembles the square wave bias function, illustrating repeatability of the activation process - the switching is reversible and repeatable, and no significant blinking or bleaching of the emitter occurs during the measurement. Similar behavior was also observed for emitters activated by a negative bias applied to the MLG.

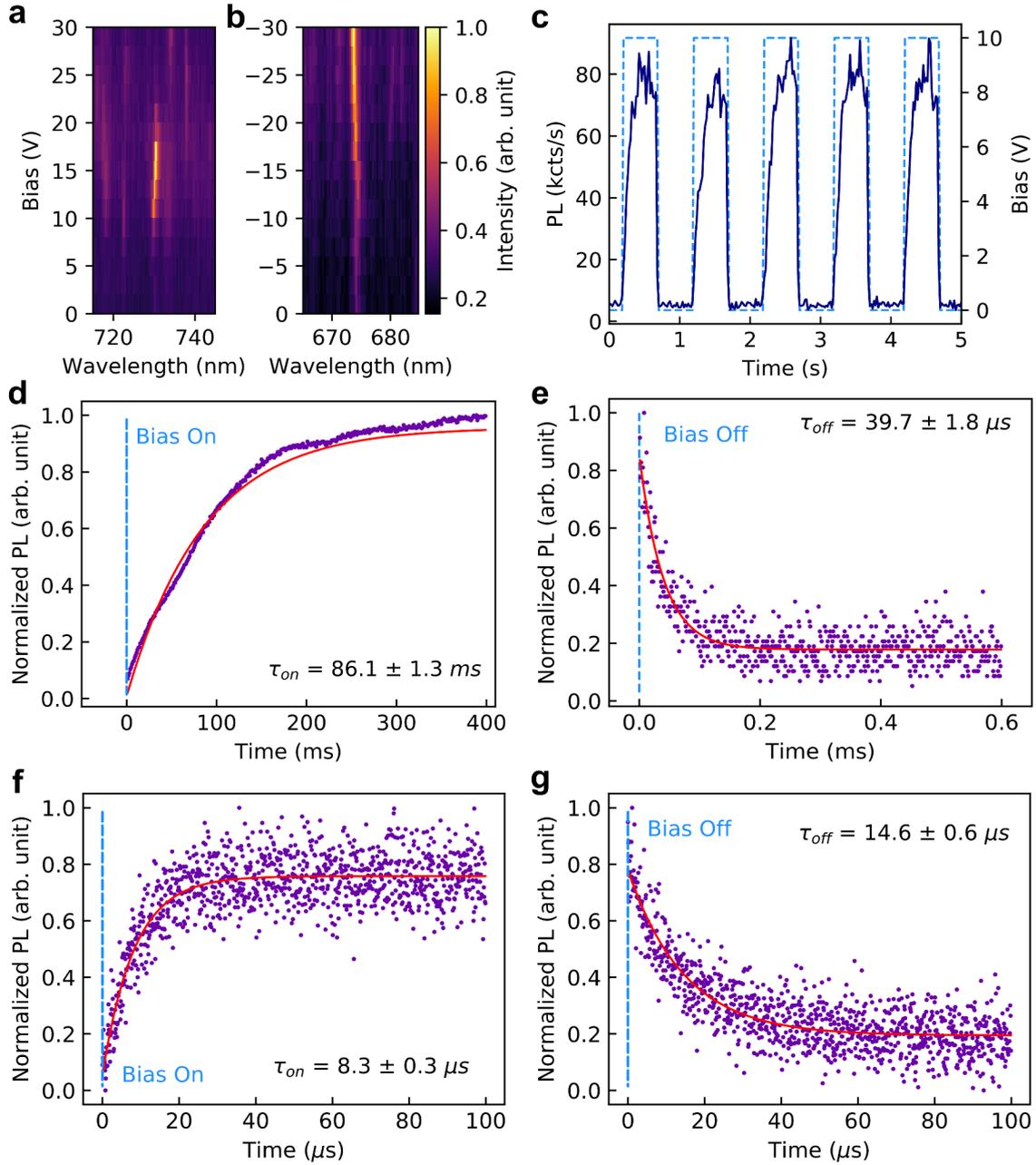

*Figure 2: Electrical control of hBN quantum emitters in the heterostructure device.* **a, b.** *Normalized PL spectra recorded from two different emitters as the bias applied to MLG was varied from 0 V to +30 V (a), and -30 V (b).* **c.** *Dynamic modulation of the emission intensity of a quantum emitter by a square wave bias function. The bias is switched periodically between 0 V and +10 V, as shown by the light blue trace. The filtered PL signal intensity detected by an avalanche photodiode (APD) is plotted in dark blue.* **d, e.** *Normalised PL intensity versus time, showing the emission dynamics when the emitter is turned on (d) and off (e) by a +10 V step function applied to the MLG electrode. The measured data are fitted with single exponential functions, and the time constants, $\tau_{on}$ and $\tau_{off}$, are 86 ms and 40 μs, respectively.* **f, g.** *Corresponding dynamics from an emitter that becomes active under negative bias, measured by applying a -10 V step function to the MLG electrode. Under negative bias $\tau_{on}$ and $\tau_{off}$ are 8 μs and 15 μs, respectively.*

A detailed analysis of the switching rates is presented in Figure 2d-g. The time-correlated intensity was recorded using a time tagger (Swabian instrument, jitter of < 200 ps) whilst bias step functions were applied to the device. Figure 2d,e show the PL rise and decay times when a bias of +10 V was turned on and off, respectively. The curves were fitted with single exponential functions and the rise ($\tau_{on}$) and fall times ($\tau_{off}$) are estimated to be ~ 86 ms and ~ 40 μs, respectively. The rise time is ~2000 times slower than the fall time, indicating significant differences between the charging and discharging dynamics[29].

The corresponding measurements obtained using a negative bias of -10 V are shown in Figure 2f, g. Under negative bias, $\tau_{on}$ and $\tau_{off}$ are comparable, approximately 8 μs and 15 μs, respectively. Strikingly, the rise time under negative bias is over four orders of magnitude faster than under the negative bias, whilst the fall times are similar under both positive and negative bias. The dramatic difference between the rise times is indicative of distinct emitter activation mechanisms under positive and negative bias, as is discussed in detail below.

To provide a broad, statistically-representative overview of the behaviour of quantum emitters under applied bias, we recorded PL spectra from a large ensemble of quantum emitters within the area of a single excitation laser spot. The spectra recorded as a function of bias over the range of -40 V to +40 V is shown in Figure 3a, where each emission line corresponds to a single quantum emitter in hBN. The lines at 580 nm (620 nm) are the G (2D) bands of MLG and remain unchanged (at this particular spectrometer resolution).

A large number of emitters spanning a broad range of emission wavelengths are activated when a positive bias is applied to the MLG electrode, mostly above +10 V. Similarly, numerous emission lines appear when a negative bias is applied to the device, and become increasingly brighter as the bias decreases to -40 V. We note that no emission was observed from the device at any bias in the absence of the excitation laser - that is, all emissions discussed in this paper are field-activated PL rather than electroluminescence.

To investigate this effect further, we plot the intensity from a number of representative emitters as a function of applied bias in Figure 3b and figure 3c. Figure 3b shows four emissions that are active within a positive bias range. The PL intensity from each of these emitters is highly bias-dependent. For example the intensity of the 581 nm line peaks at a bias of ~10V, while the 641 nm line peaks at ~28 V. Interestingly, most of the emitters have a clear bias activation range - that is, they are optically active over this range and inactive at biases outside this range. Such behaviour has never been observed for any other solid-state quantum emitters, and it is discussed in detail below.

The behaviour is substantially different when a negative bias is applied to the MLG electrode. As is shown in Figure 3a, as the bias is reduced from 0 V to -40 V, a number of emitters become optically active and none of them deactivate over the entire bias range. The intensity of a number of representative emissions from this group is plotted versus bias in Figure 3c. The emitters are very dim at zero bias, and the emission intensities increase linearly as the bias is reduced from 0 to -40 V under constant laser excitation power. We note that an increase in emitter intensity versus bias has been observed previously for neutrally charged NV centres in diamond[25, 30]. However, more broadly, the observation of PL emissions that are inactive until a voltage is applied has not been reported for any solid state quantum systems.

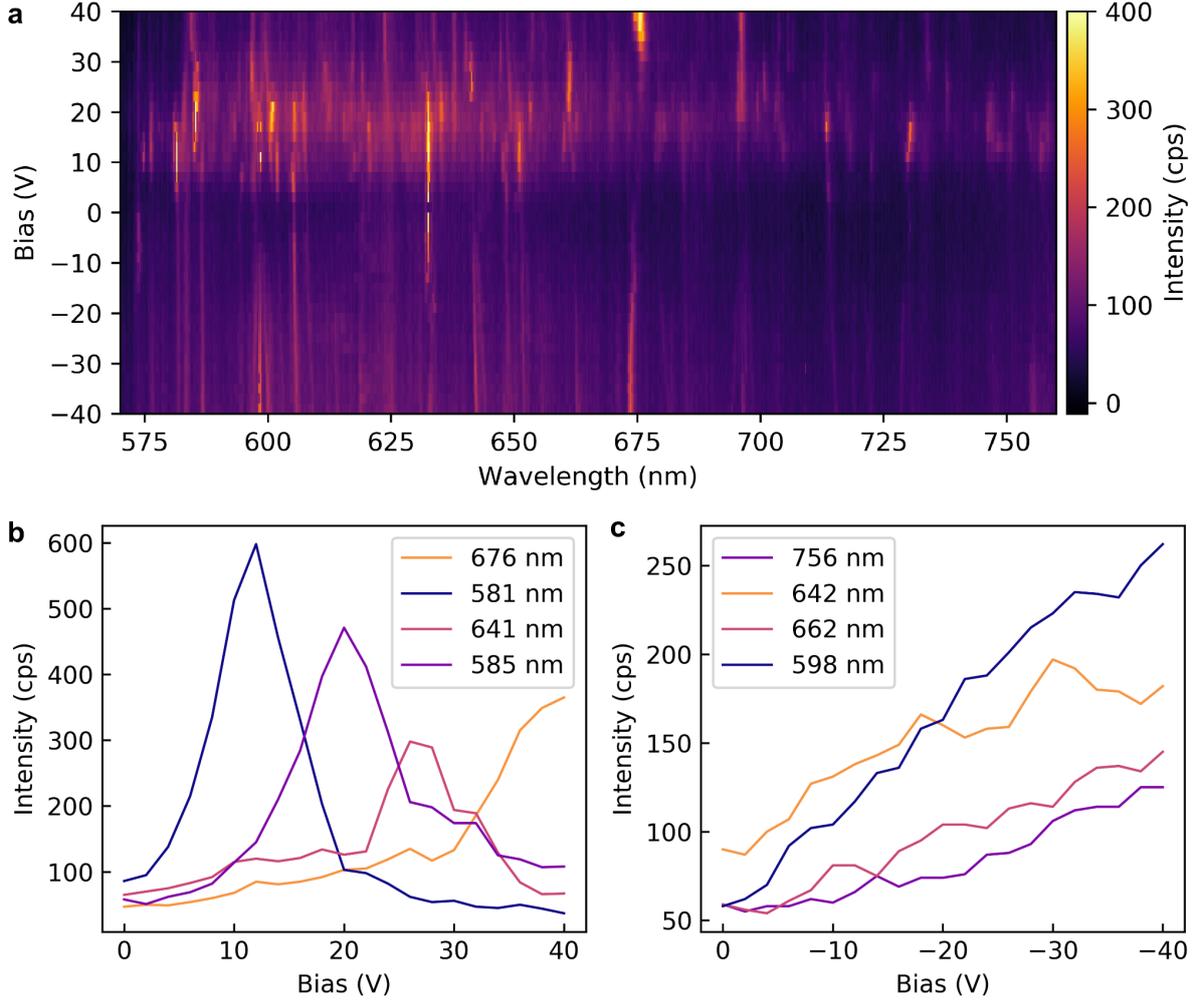

*Figure 3: Activation of hBN quantum emitters in the heterostructure device. **a.** PL spectra recorded as a function of bias, over the range of -40 to +40V. The lines at 580 nm and 620 nm are the G band and 2D band of MLG. The remaining lines are quantum emitters in hBN. **b, c** Emission intensity versus bias for a number of emitters activated by a positive (b) and a negative (c) voltage applied to the MLG electrode.*

We now turn to a discussion of the photophysics of these emitters under applied bias. We attribute the emitter activation and deactivation caused by a positive bias (seen in Figure 3b) to changes in charge states of defects in hBN, and the activation of emitters under negative bias (seen in Figure 3c) to the injection of hot electrons from MLG into hBN. These two processes are characterised by the slow and fast emitter activation dynamics, as is discussed below in the context of the electron energy level diagram shown in Figure 4.

The device band diagram under zero bias is shown in Figure 4a. The MLG split Fermi level, $E_F$, and the bottom of the hBN conduction band are located 4.6 and 2.3 eV below the vacuum level, respectively. Also shown on the diagram are two hypothetical charge transition levels of a defect in hBN, adapted from reported density functional theory (DFT) calculations[31, 32]. Figures 4b and c show the device at a bias of +10 and +20 V, respectively, and illustrate how a positive bias sweep causes sloping of the energy bands, and an effective sweep of $E_F$ within a subset of the band gap of hBN. A defect with a charge transition level within this region of the band gap will gain/lose an electron as $E_F$ moves above/below the level (Figure 4 b).

Similarly, a defect with two charge transition levels in this region of the band gap will change charge state twice if $E_F$ sweeps through both levels. Hence, the hBN defect in Figure 4a will have lost two electrons upon the application of + 20 V to the MLG (Figure 4c). Each change in the charge state of an emitter will result in a corresponding change in the defect energy levels and hence the emission spectrum[25, 30]. Importantly, a change in charge state often causes activation or deactivation of an emitter – either absolutely or effectively by causing the emission energy to shift outside the measured spectral range[25]. Hence, activation of a hBN quantum emitter upon the application of a positive bias to the MLG electrode of our heterostructure device can be caused by a change in the charge state of the emitter by +1 (Figure 4b). Deactivation of the emitter at a greater positive bias can be caused by a second change in charge state, provided that $E_F$ crosses a second charge transition level of the emitter (Figure 4c). For completeness, we should point out that the voltage sweeps also affect defect states that are in the vicinity of and can modulate quantum emitters in hBN.

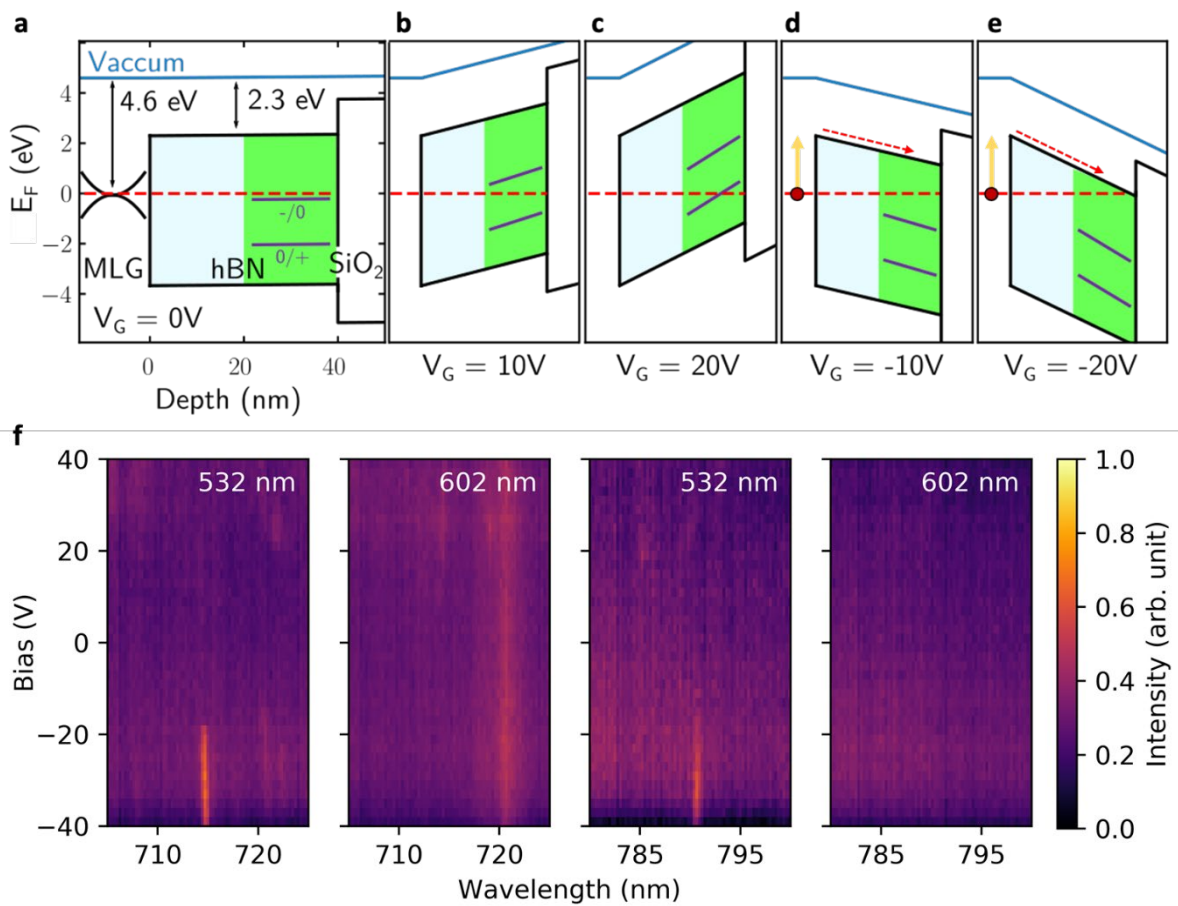

*Figure 4: Band diagram of the heterostructure device under various bias configurations. a. The device with both electrodes grounded ($V_G = 0V$). The hBN capping layer is shown in light blue and the hBN layer that contains quantum emitters is shown in green (the thickness of each hBN layer is assumed to be 20 nm). The MLG split fermi level ($E_F$) extends into the hBN, indicating charge transfer between MLG and defect states in hBN (see text). The purple lines indicate two hypothetical charge transition levels of a single defect in hBN. b,c. The device with a bias of +10 V (b) and +20 V (c) applied to the MLG electrode. d,e. The device with a bias of -10 V (d) and -20 V (e) applied to the MLG electrode. The solid yellow arrows show*

*photoexcitation of an electron in MLG, and broken red arrows indicate drift of the electrons via the band gaps of the hBN layers. **f.** Experimental verification of the model, whereby emitters are only visible under green excitation (at ~ 715 nm and 790 nm), but not under lower energy red excitation, at the same confocal spot. The emission at 720 nm is the graphene Raman.*

Based on the above, activation of an emitter upon application of a negative bias could be argued to be caused by a change in the charge state of the emitter by -1. However, an upward sweep of $E_F$ within the bandgap of hBN will populate deep defect levels and we do not expect it to activate emitters. Moreover, we found that the activation rate measured by applying a step voltage function to the device is over three orders of magnitude slower for the case of positive bias than for the case of negative bias (Figure 2d and f, respectively), indicating a fundamental difference in the charge transfer dynamics. To explain this difference, we consider energy band diagrams for the negatively charged device shown in Figure 4d and e for the case of -10 and -20 V, respectively. Application of a bias that is negative with respect to the MGL electrode inverts the gradient of the sloped bands and effectively raises $E_F$ towards the hBN conduction band. In this configuration, electrons excited in the MLG by the laser (yellow arrows in Figure 4) can tunnel across the barrier at the MLG-hBN interface and drift (red broken arrows) within hBN under the influence of the applied electric field. The resulting photocurrent provides a means to supply hot electrons to emitters *via* the hBN conduction band. This charge transfer mechanism is therefore expected to be fast relative to the case of a positive bias (Figure 4b,c), where electron removal from the deep hBN charge transition levels likely occurs *via* a hopping mechanism and electrons flow to the MLG *via* trap states inside the hBN band gap.

The above analysis illustrates two distinct charge transfer mechanisms between the MLG electrode and defects in hBN, which are slow/fast in the case of positive/negative bias applied to the MLG. The first can account for emitter activation and deactivation upon application of a positive voltage sweep to the device, and the second can account for emitter activation by a negative bias. We note that the almost universal deactivation of emitters at +40V seen in Figure 4a is likely a consequence of the fact that $E_F$ lies very close to the hBN valence band and the ground states of most emitters are ionised at this voltage. We also note that, as is evident from Figure 4, the voltage needed to activate/deactivate various emitters is a function of the emitter location within the hBN. This observation combined with the fact that a number of distinct defect species are responsible for the rich emission spectrum of hBN accounts for the variation in activation and deactivation voltages seen in Figure 3a.

To provide further experimental support for our model, we increased the excitation laser wavelength from 532 nm (~2.3 eV) to 602 nm (~ 2 eV). The longer wavelength excitation should not be sufficient to overcome the energy barrier (see figure 4 d,e) under negative bias, and hence no emitters should be activated. Indeed, this hypothesis is confirmed. Figure 4f shows PL spectra of emitters under positive and negative bias recorded at the same confocal spot using the two excitation wavelengths. It is clear that new emitters appear under negative bias when a 532 nm excitation laser is used, but no emission appears under the longer excitation wavelength of 602 nm.

To illustrate the potential of our devices for practical and scalable quantum photonic applications, we demonstrate resonant excitation of these quantum emitters under a negative bias. We expect that under these conditions, the electric field and photocurrent will govern the charge states of both emitters and surrounding charge traps and thus suppress charge fluctuations and spectral diffusion. Furthermore, the photocurrent is expected to re-activate emitters after each cycle, without the need for a repumping laser, as is common for most other solid state defects, including defects in hBN. This was indeed observed, as is illustrated in figure 5. Figure 5a shows an emitter with a ZPL at ~ 588.5 nm, recorded from the device using an off-resonant 532 nm excitation laser. The off-resonant linewidth is phonon broadened as expected. Figure 5b shows a resonant excitation scan of the same emitter with a measured linewidth of ~ 158 ±19 MHz. Both measurements were taken using an applied bias, $V_G$= -40V, and importantly, no resonant emission was observed at zero bias. For quantum emitters in hBN with excited state lifetimes on the order of ~ 3 ns, ~ 160 MHz certainly represents a nearly-coherent, Fourier Transform limited, linewidth, which is highly promising for future-generation indistinguishable photons.

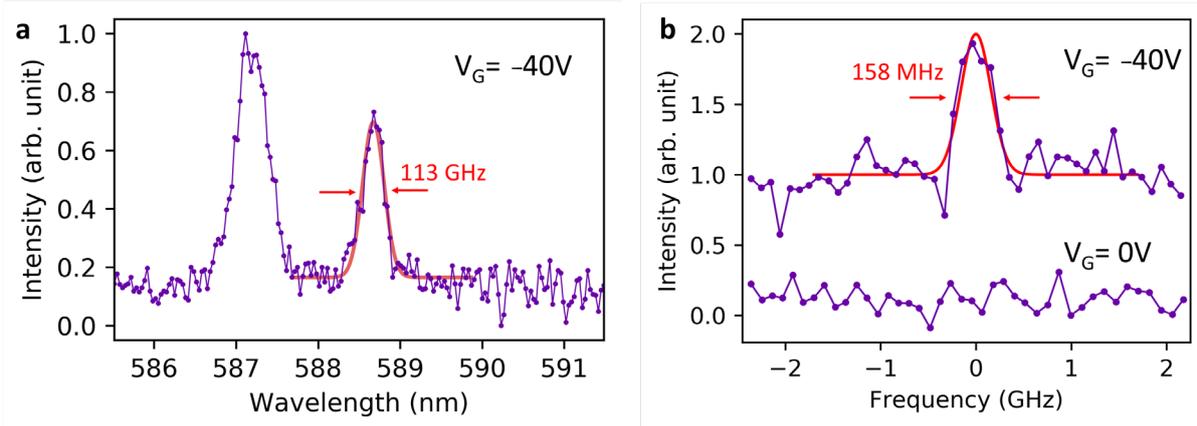

*Figure 5. Coherent excitation of quantum emitters in hBN. a.* Emission spectrum of a single emitter with a ZPL at ~ 588.5 nm, recorded under non-resonant 532 nm excitation. *b.* Resonant excitation of the same emitter, resulting in a nearly-coherent photon source with a linewidth of ~ 158 MHz. Both measurements were done using a bias voltage, $V_G$= –40V.

To summarise, we demonstrate electrical modulation and control of a variety of quantum emitters in a vdW heterostructure. The quantum opto-electronic devices consist of MLG/hBN heterostructures, operate at accessible voltages and can be assembled using readily-accessible fabrication techniques. We propose two distinct mechanisms for device operation versus bias polarity based on electrostatic charge switching of quantum emitters and drift of hot photoelectrons. Our results open a plethora of new opportunities in integrated quantum photonics with vdW materials. First, the ability to modulate and switch on/off quantum emitters is imperative for scalable quantum circuitry. Second, electrostatic gating can now be used to activate emitters post hBN growth and processing, and to select emitters at specific wavelengths. Third, a single device can now be employed to activate and tune emitters into resonance to achieve indistinguishable photons from quantum emitters in hBN. Indeed, our results already show that under negative bias a nearly-coherent quantum source in hBN with linewidths of ~ 160 MHz can be obtained. Finally, and equally important, our results constitute the possibility to characterise charge transition levels of specific defects in hBN, and correlate them with theoretical studies of specific atomic defect structures.

## Acknowledgements
The authors acknowledge financial support from the Australian Research Council (CE200100010, DP190101058) and the Asian Office of Aerospace Research & Development (FA2386-20-1-4014). The authors thank the Australian Nanofabrication Facilities at the UTS OptoFab node.

## Author Contributions
I.A., Z.X., M.K., and S.W. conceived the idea and designed experiments. T.Y. carried out the device fabrication, C.L., and Z.X. aided in emitter engineering and S.W. carried out the cryogenic measurements. S. W., M. K., T.Y. plotted and analysed the data with input from all coauthors. N.D. and A.S. performed theoretical simulations. I.A. M. T. and M. K. supervised the work and wrote the manuscript with input from all coauthors.